\documentclass[letterpaper, 10 pt, conference]{ieeeconf}  

\IEEEoverridecommandlockouts                              

\usepackage{booktabs} 
\usepackage{footnote}
\usepackage[flushleft]{threeparttable}
\usepackage{amsmath}
\usepackage{graphicx}
\usepackage{url}
\usepackage{color}
\usepackage{float}
\usepackage{flushend}
\usepackage{textcomp}
\usepackage{gensymb}
\usepackage{subcaption}
\usepackage{wrapfig}
\usepackage[symbol]{footmisc}
\usepackage[noadjust]{cite}
  
\IEEEoverridecommandlockouts

\DeclareMathOperator*{\argmin}{arg\,min}

\title{\LARGE \bf
The Interface Usage Skills Test: An Open Source Tool for Quantitative Evaluation in Real Time for Clinicians and Researchers
}

\author{Mahdieh Nejati Javaremi$^{1,4,5}$, Sisilia Sinaga$^{2}$, Yuming Jin$^{1}$, Matthew L. Elwin$^{1}$, and Brenna Argall$^{1,2,3,4,5}$ 
\thanks{$^{1}$Department of Mechanical Engineering, Northwestern University, USA}%
\thanks{$^{2}$Department of Computer Science, Northwestern University, USA}%
\thanks{$^{3}$Department of Physical Medicine and Rehabilitation, Northwestern University, USA}%
\thanks{$^{4}$Shirley Ryan AbilityLab, Chicago, IL 60611, USA}%
\thanks{$^{5}$Corresponding authors {\tt\small m.nejati@u.northwestern.edu}, {\tt\small brenna.argall@northwestern.edu}}
}

\begin{document}

\maketitle
\thispagestyle{empty}
\pagestyle{empty}

\begin{abstract}

Assistive machines endow people with limited mobility the opportunity to live more independently. However, operating these machines poses risks to the safety of the human operator as well as the surrounding environment. Thus, proper user training is an essential step towards independent control and use of functionally assistive machines. The human operator can use a variety of control interfaces to issue control signals to the device, depending on the residual mobility and level of injury of the human operator. Proficiency in operating the interface of choice precedes the skill in operating the assistive machine. In this systems paper, we present an open source tool for automatically and objectively quantifying user skill in operating various interface devices. 

\end{abstract}

\section{INTRODUCTION}

Assistive machines enable people with motor impairments or other mobility deficits to achieve a higher level of independence, community and social participance, and quality of life. Although these devices expand a person's capabilities, if handled without proper training, they can also pose safety hazards to the human using the device as well as those around them. Thus, many individuals who can benefit from an assistive machine such as a powered wheelchair are barred from using one because of inadequate control proficiency~\cite{Kairy2014}. 

\textit{Proficiency in handling the interfacing device precedes proficiency in handling the assistive machine}. The prescription of the most appropriate interface device to an individual based on their preferences and capabilities, the adjustment of the interface to the specific needs and abilities of that person, and the subsequent training of them in the safe and skilled use of that interface are integral elements of the preparation of the human to operate an assistive machine with greater success. For both clinicians and researchers to develop better-informed strategies and technologies that overcome operational challenges still faced by current and potential powered wheelchair (PW) users, it is imperative to have a standard tool for quantitative assessment of interface usage skill. However, there is a lack of \emph{objective} quantitative measurements of \emph{interface} use. Considering the specific case of a PW, there is a lack of objective quantitative assessment of the wheelchair user's navigational control skills in each step of the process of gaining access to and using a PW---from introducing the individual to a PW by a wheelchair professional, the selection of an appropriate interface in a wheelchair seating clinic, and ongoing training with a physical therapist. Clinicians use qualitative and subjective observations and their previous experience as a way to gauge which settings and interfaces are suitable for the patient. Therapists commonly use the Wheelchair Skills Test (WST)~\cite{wstwebsite}, the current clinical standard for measuring wheelchair skills, to evaluate a patient's performance by assigning a discrete capacity score, again through \emph{subjective observations}. The System Usability Scale (SUS) is another commonly used tool that is completed by a user to assess the subjective usability of various products~\cite{bangor2008empirical}. However, this tool does not provide a quantitative measure of the user's skill in using the specific product. Access to analytics of the person's interface usage skill can help identify areas of deficit in order for clinicians to provide better informed and targeted training and therapy for the ultimate goal of improving functional assistive machine usage.  

There is currently no standard assessment tool for evaluating a user's ability to control an interface for driving an assistive device. We fill this gap with the work described in this systems paper. We present a real-time analytics suite with conversion to app form suitable to run on any computer platform and a free beta release on the Android Play Store\footnote{Source code: https://github.com/argallab/InterfaceUsageSkillsTest.git}. The tool can be used by clinicians without additional training and provides on-the-fly statistics on multiple interface usage measures. Raw data are also stored, which can be used for a more detailed analysis of custom interface usage characteristics. We also contribute hardware specifications for an adaptor that allows common commercial interfaces used for assistive machines to communicate over Bluetooth with our assessment tool.


First, we cover a brief background on the relevant literature in Section~\ref{sec:background}. We then provide a detailed description of our Interface Skills Test software and hardware evaluation tools in Sections~\ref{sec:app} and~\ref{sec:hardware}, with a discussion of the preliminary results. Implications for clinical and research use is covered in Section~\ref{sec:discussion}. We conclude with our proposal for future work in Section~\ref{sec:conclusion}.

\section{Background}\label{sec:background}
In this section, we present a summary of related work on assistive machine skills assessment tools and characterizing interface usage. 

\subsection{Powered Wheelchair Skill Measures}
In the domain of functional rehabilitation, outcome measures span a wide range from kinesthetic and neurophysiological~\cite{boninger1999wheelchair} measures of patients to global outcomes in terms of overall function and community reintegration~\cite{wood1988assessment}. 

The Wheelchair Skills Test (WST) is the state-of-the-art in clinical evaluation of an individual's ability to drive a PW, and is an intermediate-level outcome that lies between the two extremes cited above~\cite{Kirby2002}. This measure consists of various tasks---including ascending and descending slopes, and navigating through doorways---and for each task, an observing therapist chooses a capacity and confidence based on the completion of the task and subjective safety. A score of 0 indicates failure to complete the task, 1 indicates that the user had some difficulties completing the task, and 2 indicates that they accomplished the task without any difficulty. This measure does not capture details of the exact difficulties the person experienced in completing specific tasks, such as control smoothness. Although the WST is a powerful assessment tool, its delivery is subject to clinician training~\cite{giesbrecht2022wheelchair}. 
Powered Mobility Clinical Driving Assessment (PMCDA) is another assessment tool that is also observational~\cite{Candiotti2019}. 

Currently, there are no assessments that consider how an individual completes the skill in terms of objective measures of safety, such as distance to barriers, speed, or smoothness profiles~\cite{routhier2003mobility}. Furthermore, in the assistive robotics domain, there is no standard way to assess user skill in operating a teleoperaton interface. The most common performance measures used to evaluate the efficacy of assistive robotic systems include task completion time and tracking errors, or subjective questionnaires that do not assess objective user skill~\cite{erdogan2017effect, Olsen2003, Goodrich2004}.

All of these assessments require a specific physical setup and, more importantly, do not provide a quantitative analysis of \textit{interface usage}, but rather qualitative observations on wheelchair driving skill. We propose that by looking at one level of abstraction---the interface operation---we can better identify areas of control deficit that will lead to improving overall wheelchair driving skill. 

\subsection{Interface Usage Characterization}

To our knowledge, there is no standard assessment tool for characterizing interface usage. Research studies of interface characterization have focused mainly on the neuromuscular and physical response of the human during manual control to study human sensing and response characteristics~\cite{sheridan1974man,kleinman1971control,burchfield1967optimal}. 

In the field of assistive technology, clinicians have been surveyed on the usefulness and adequacy of powered wheelchair control interfaces for their patients~\cite{Fehr2000_interfaces}. Their results provide subjective evidence regarding steering and maneuvering difficulty, and for a need to integrate robot autonomy into conventional powered wheelchairs. However, their results do not provide quantitative information on interface usage, which could be exploited by assistive autonomy. Novel interface technologies have been developed for those with severe motor impairments, including an isometric joystick~\cite{Cooper2002a_isojoy}. The authors compared task completion time and accuracy of the novel interface to a conventional hand-held joystick within a control population but not an end-user population. In another work, the authors introduce a novel interface and compare user precision and performance between the target spinal cord injured (SCI) group and an uninjured control group, but do not compare the performance against commercially available interfaces~\cite{Farshchiansadegh2014_bmi}. 

In the human-robot interaction domain, a study investigated the influence of video game usage on human-robot team performance~\cite{richer2006video}. The authors classify interfaces based on their inputs and outputs and use the information to create a framework for systematically evaluating interfaces in the Human Robot Interaction (HRI) domain. Prior work in remote teleoperation has also studied the effect of time delay and communication channel degradation on the quality of teleoperation and manual control~\cite{timedelay_space_survery}. 


In this work, we contribute an assessment tool for the quantitative characterization of interface usage.

\section{Interface Skills Test}\label{sec:app}
In prior work, we introduced a variety of interface usage metrics to characterize interface usage skill~\cite{icorr2019}. In this work, we expand upon these measures and design an open source application that 
computes these measures online in real time and stores data within a user profile that is easy to use by anyone familiar with a smart phone, tablet, or computer. No training is needed to use this tool, and analytical graphs are available for the clinician and researcher, as well as the human operator of the assistive machine. In this section, we introduce the interface skills test, describe key variables that impact interface usage, configurable settings, and outcome measures. 

\subsection{Tasks}
The assessment consists of a series of tasks that evaluate qualities of the human input including speed, precision, and stability: all necessary criteria for a human to be able to properly command an assistive machine.
This is done through two distinct tasks: 
a command following task and trajectory following task. 
Tasks are designed in a simulated environment so that uncertainties from real world dynamics do not corrupt the interface usage performance measures.

Individual user profiles can be made and the tasks can be selected via a menu (Fig.~\ref{fig:menu}). Each task can be reconfigured as described in Section~\ref{subsec:settings}. 

\subsubsection{Command Following} \label{sec:command}
The command following task is designed to uncover a patient's ability to respond to a visual command stimulus in terms of response accuracy, speed, and stability (Fig.~\ref{fig:command_following1}). In this task, a white arrow---the command prompt $\hat{\textbf{u}}$---appears on the screen pointing in different directions in a random and balanced sequence. The default direction settings include the four cardinal and four inter-cardinal angles. The length of the arrow also changes in a random and balanced order to measure how well the human can scale inputs to the prompted command. The human is instructed to issue a command for the same direction and magnitude (if the interfacing device allows for scaling) as soon as they see the command prompt and to continue issuing the command uninterrupted for the duration of the prompt ($T$). The blue arrow is the feedback of the actual command issued by the human.

\subsubsection{Trajectory Following} \label{sec:traj}
\begin{figure}[t]
\centering
\begin{subfigure}[thb]{.22\textwidth}
   \includegraphics[width=\textwidth]{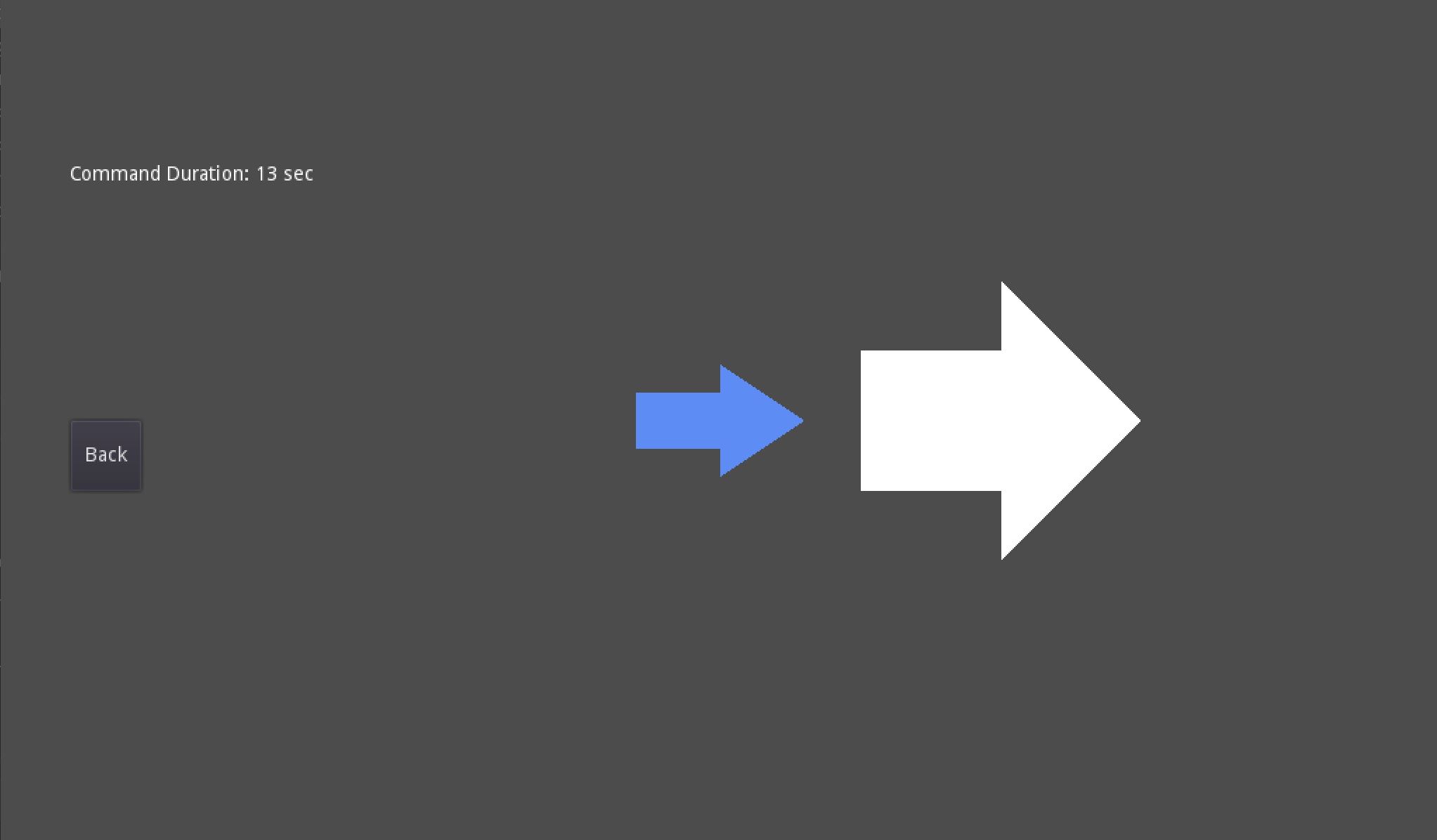}
    \caption{}
    \label{fig:command_following1}
\end{subfigure}
\begin{subfigure}[thb]{.2\textwidth}
   \includegraphics[width=\textwidth]{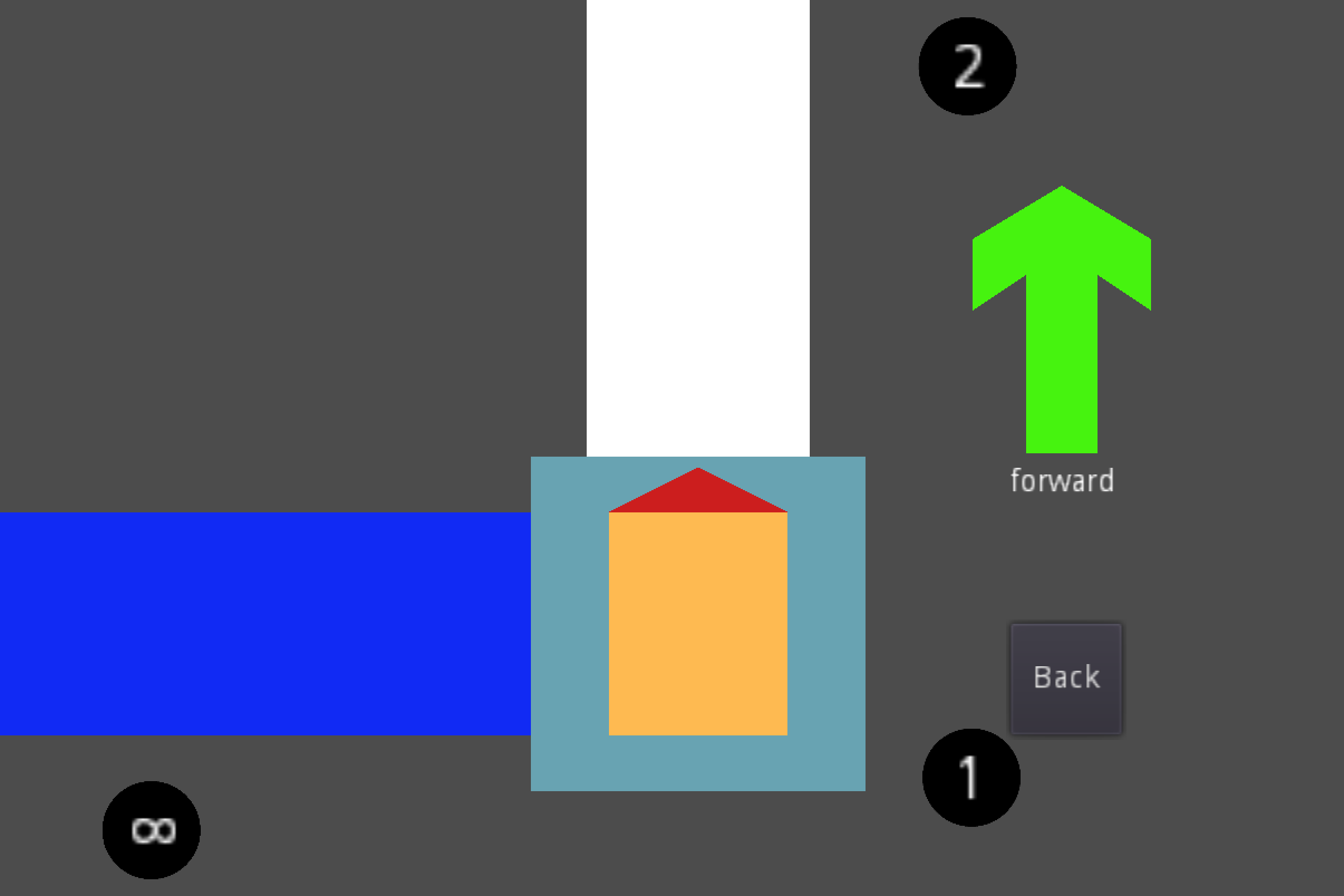}
    \caption{}
    \label{fig:traj_square}
\end{subfigure}
\caption{Study tasks. (a) Command following task. Target prompt (white arrow) and human response (blue arrow). (b) Trajectory following task. Direction of motion prompt (green arrow), wheelchair footprint (yellow square and red front), and path segments (white and blue, curves and lines).
}
\vspace{-0.5cm}
\label{fig:tasks}
\end{figure}

The trajectory following task is designed to evaluate how well the human is able to follow a predefined path to evaluate signal integrity in terms of smoothness, ability to give corrections, and directness of the human command. Trajectory following can be thought of as the inherent ability to generate commands to follow waypoints while using visual feedback correction. The ability to follow a trajectory where there is a single known goal---without interference from the wheelchair dynamics and external sources of noise---aims to uncover how a person's intended goal may differ from the signal they output through the interface. The task consists of controlling the motion of a 2D simulated wheelchair (the yellow and red pentagon shape in Fig.~\ref{fig:traj_square} and Fig.~\ref{fig:traj_curve}) along a predefined path. The path is demarcated to indicate motion along goal posts. The trajectory path begins with a square path, followed by a curved path. Only the path in the immediate vicinity of the wheelchair is visible at any given moment (as in Fig.~\ref{fig:traj_square}). The patient is instructed to stay within the bounds of the clearly marked path and to avoid going into the out-of-bounds grey area. The square and curved paths are designed such that they contain the basic commands covered by general interfacing devices used for 2D assistive machines. The square path consisted of two forward, two backward, two $90\degree$ left turn and two $90\degree$ right turn trajectories. The curved path consists of two long arcs and two small arcs.

\subsection{Key Variables}
A multitude of variables impact the human's interface usage characteristics while operating an assistive machine. We group these variables into four distinct categories~\cite{mcruer1967review}: 
\begin{itemize}
 \item \textbf{Operational Variables.} 
 Factors such as the dynamics of the device they are controlling (e.g., rear-wheel vs mid-wheel drive wheelchair), the mechanics of the control interface, as well as what information is available to the human and how. 
 \item \textbf{Environmental Variables.} Factors such as temperature, whether the task is indoors or outdoors, or noise.
 \item \textbf{Internal Variables.} Factors such as internal motivation, training and skill, fatigue, and stress that affect human internal states.
 \item \textbf{Procedural Variables.} Features of the experimental design, such as the instructions for the given task and order of task presentation.
\end{itemize}
 
The effect of procedural variables is minimized in the Interface Skills Test by standardizing the instructions for each task within the app. To control for the remaining variables, the clinician, test taker, or test giver can input information using questionnaires prior to and after a given task, as described in the following section.  

\subsection{Configurable Settings}\label{subsec:settings}
The assessment tool is designed with a variety of configurable input settings through a user-friendly GUI. The configurable measures consist of information relating to operational variables, environmental variables, and the human's internal variables, as well as configurable settings pertaining to how the tasks are presented to the human test taker. 

The controlled covariates for documenting the human's internal state are collected through a series of Likert-type questionnaires administered directly within the app. Documenting these variables is not necessary for running the assessment, but it is important and recommended to keep track of these as they may inform larger trends in the human's interface usage skill characteristics. 

\begin{itemize}
    \item \textbf{Fatigue}: Measured using the Fatigue Scale, which is an 11-item Likert-type questionnaire~\cite{chalder1993development}. 
    \item \textbf{Motivation}: Measured using the Intrinsic Motivation Inventory (IMI)~\cite{imi}. 
    \item \textbf{Workload}: Measured using the raw NASA-TLX which is a shortened version of this assessment tool~\cite{tlx}. 
    \item \textbf{Stimulant consumption}: Text entry question. 
    \item \textbf{Confidence}: A 5-point rating scale question on how confident the human is in their ability to use the interfacing device. 
    \item \textbf{Stress}: Measured via the Perceived Stress Questionnaire~\cite{fliege2005perceived}.
\end{itemize}

Other control variables can also be documented that monitor environmental and operational variables: 
\begin{itemize}
    \item The interface used during the test. 
    \item How often the interface is used daily by the patient. 
\end{itemize}

\begin{figure*}[t]
\centering
\begin{subfigure}[thb]{.17\textwidth}
    \includegraphics[width=\textwidth]{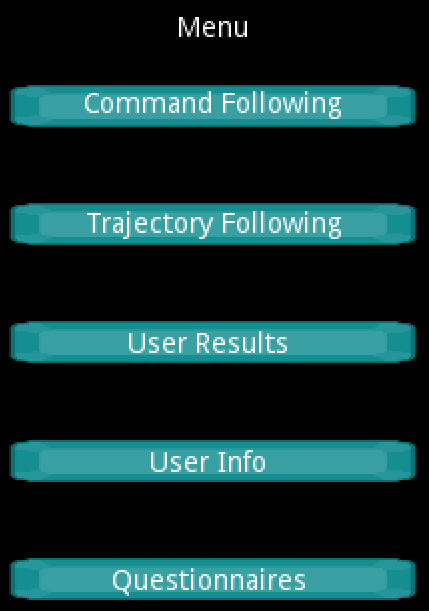}
    \caption{}
    \label{fig:menu}
\end{subfigure}
\begin{subfigure}[thb]{.19\textwidth}
    \includegraphics[width=\textwidth]{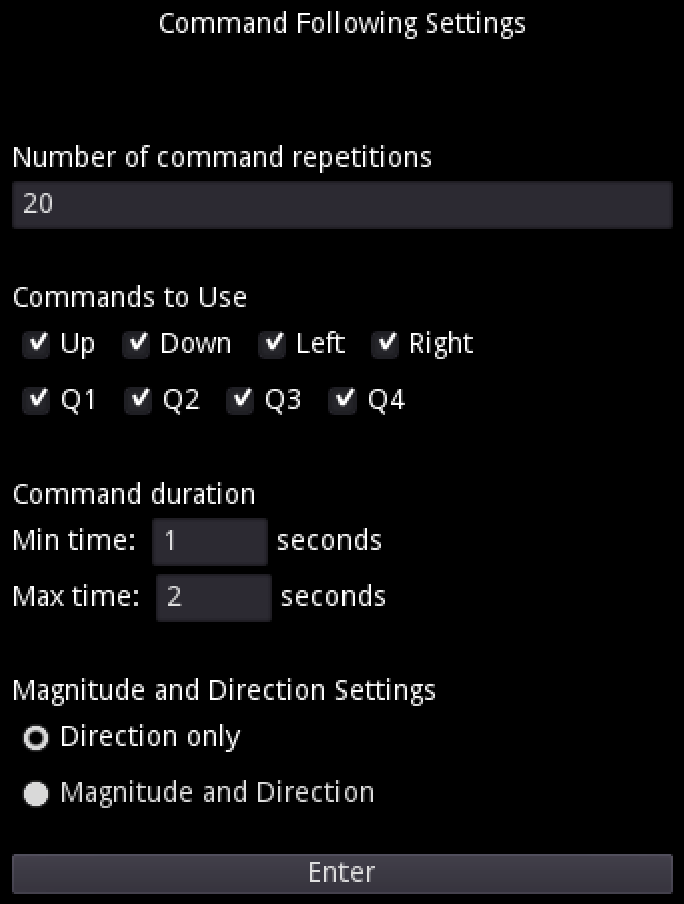}
    \caption{}
    \label{fig:menu_comm}
\end{subfigure}
\begin{subfigure}[thb]{.20\textwidth}
   \includegraphics[width=\textwidth]{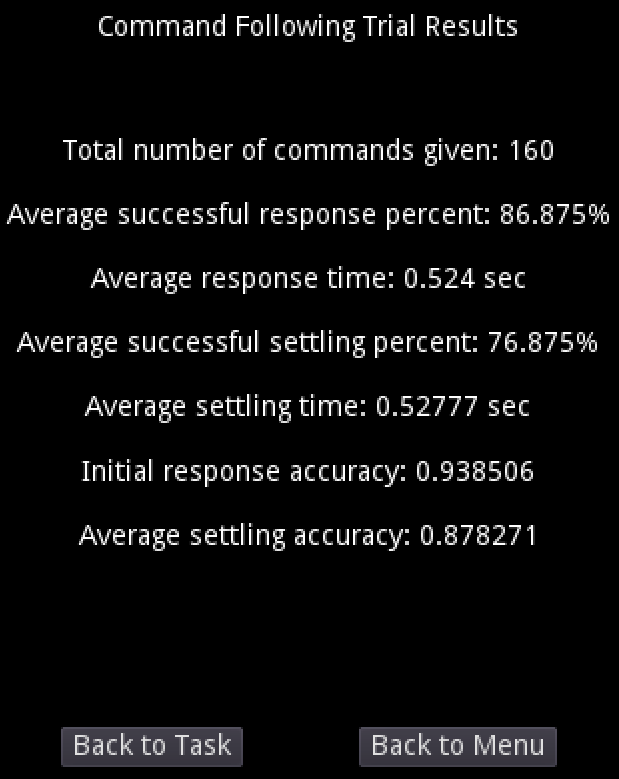}
    \caption{}
    \label{fig:summary_stats}
\end{subfigure}
\begin{subfigure}[thb]{.19\textwidth}
  \includegraphics[width=\textwidth]{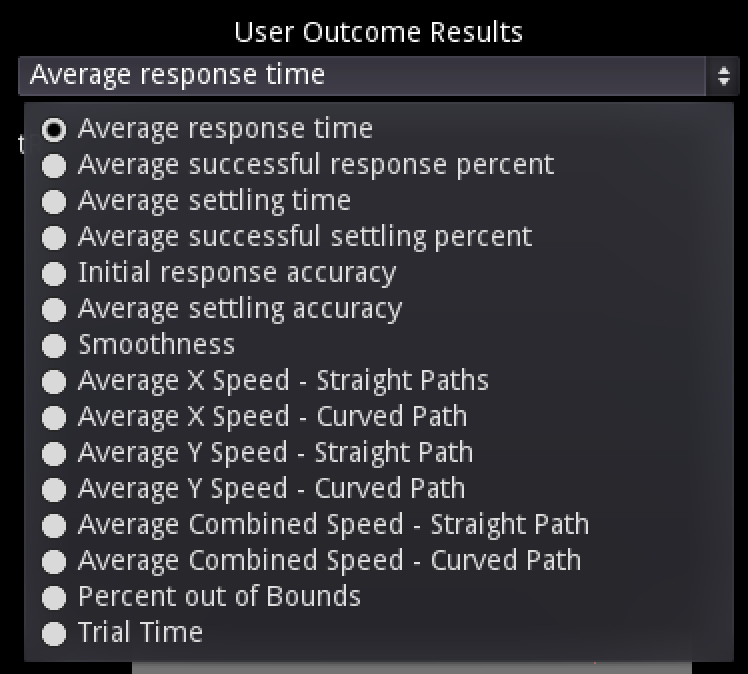}
    \caption{}
    \label{fig:results_menu}
\end{subfigure}
\begin{subfigure}[thb]{.18\textwidth}
   \includegraphics[width=\textwidth]{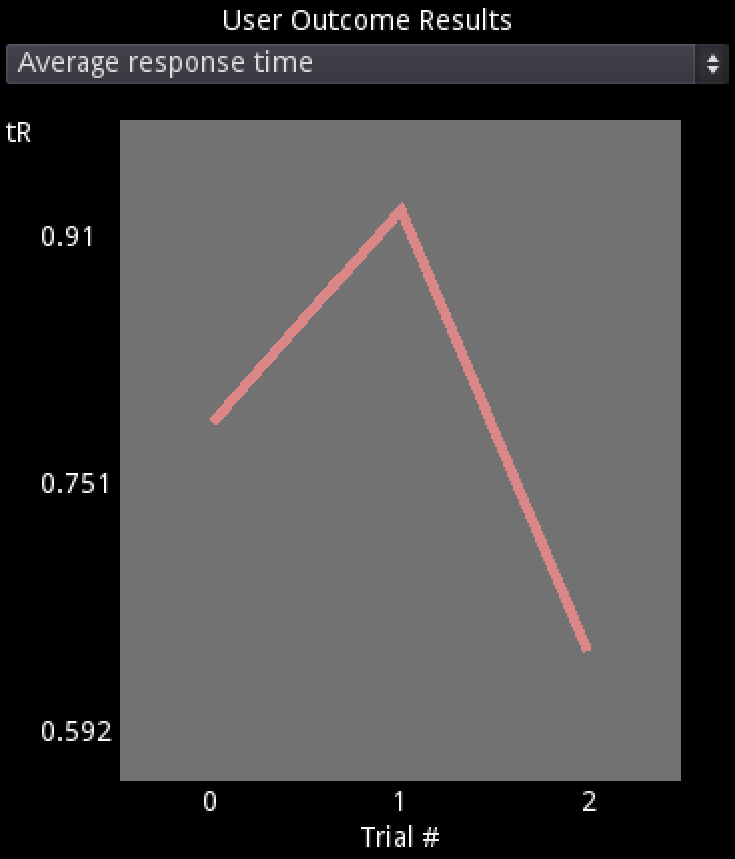}
    \caption{}
    \label{fig:results_plot}
\end{subfigure}
\caption{(a) Interface Skills Test assessment main menu. (b) Command following task configurable settings menu. (c) Command following summary statistics available immediately after trial completion. (d) Drop-down menu of available outcome measure results. (e) Plot of a single outcome measure for a particular patient's result over multiple trials.}
\label{fig:results}
\end{figure*}

Independent variables used within the command following tasks able to be reconfigured by the clinician and test-taker include (Fig.~\ref{fig:menu_comm}): 
\begin{itemize}
    \item Set of target control commands. This also includes the choice of selecting only directions or magnitudes of the commands. The default is the four cardinal and four inter-cardinal angles. 
    \item Number of times each target command is prompted. The default is set to 20. 
    \item Range of time each target prompt is displayed. This is the amount of time each command prompt is visible and the time the user has to respond. The default range is between 1-2 seconds.
\end{itemize}


\subsection{Outcome Measures and Scoring}
One of the main contributions of our work is that the assessment is calculated using \underline{strict closed-form equations}, and that the results do not depend on qualitative observations of the test taker. Additionally, outcome measures are calculated while the assessment is being administered, and \underline{results are available immediately} following the conclusion of the assessment.  A summary of the performance statistics is available immediately after the conclusion of a test trial, accessed via the GUI as seen in Figure~\ref{fig:results}.

The outcome measures available immediately after the command following task include:

\begin{itemize}
    \item \textit{Average response delay:} The average of all time differences between target command prompts and the first instance when the patient issues the correct command. 
    \begin{equation*}
        \centering
        \begin{split}
           \qquad \qquad \qquad t_d &= \frac{1}{M} \sum_{m=1}^{M} \tilde{t}^m_* \\
            \text{where} \quad \tilde{t}^m_* = & \argmin_{t\in(0, T^m]} \{ ~\big|\textbf{u}^t_h -\hat{\textbf{u}}^m|< \epsilon~ \}
        \end{split}
    \end{equation*}
    $M$ is the total number of target command prompts, and $\hat{\textbf{u}}^m$ is the $m^{th}$ target command prompt which lasts for a duration of $T^m$. $\textbf{u}^t_h$ is the patient command at time $t$, and $\tilde{t}^m_*$ the time when this command first comes within tolerance $\epsilon\pm5\degree$ of the target prompt. 
    The clock restarts for each target command prompt.\footnote{When the dimensionality of $\hat{\textbf{u}}$ is greater than 1, the $arctan(\textbf{u}^t_h, \hat{\textbf{u}}^m)$ operator is used to compute the difference $~\big|\textbf{u}^t_h -\hat{\textbf{u}}^m|$.}
    \item \textit{Average successful response percent:} The percentage of command prompts to which the patient is able to respond successfully. 
    \begin{equation*}
        \begin{split}
            &\qquad \qquad \qquad r_p = 100 \cdot \frac{1}{M}{\sum_{m=1}^{M} I^{m}}\\
            I^m &=\begin{cases}
            1, & \text{if $\exists \enskip t\in(0, T^m] \enskip s.t.~ |\textbf{u}^t_h-\hat{\textbf{u}}^m|<\epsilon$}\\
            0, & \text{otherwise}.
            \end{cases} 
        \end{split}
    \end{equation*}
   where $I^m$ is a tracking index. 
    \item \textit{Average settling time:} The time it takes until the patient continually issues the prompted command within an allowable tolerance of $\epsilon$. 
    \begin{equation*}
    \begin{split}
        &  \qquad \qquad t_s =  \frac{1}{M} \sum_{m=1}^{M} t^m_*  \\
        \text{where} ~~~t^m_* &=\argmin_{ t\in(0,T^m]} 
        \{ ~\big| \textbf{u}^k_h-\hat{\textbf{u}}^m|<\epsilon ~ \forall ~ k\in[t,T^m]~\}
    \end{split}
\end{equation*}
$t^m_*$ is the time from which the human command remains within tolerance.
    \item \textit{Initial response accuracy:} How close on average the first within-tolerance 
    response is to the target prompt.
    \begin{equation*}
        a_{r} = \frac{1}{M} \sum_{m=1}^{M} [1-|\tilde{\textbf{u}}^m_* - \hat{\textbf{u}}^m|]
    \end{equation*}
     \text{where} $\tilde{\textbf{u}}^m_*$ is the patient command at time $\tilde{t}^m_*$. 
    
    \item \textit{Average settled accuracy:} How close on average the within-tolerance response
    is to the target command after having settled.
    \begin{equation*}
        a_{s} = \frac{1}{M \cdot (T^m - t^m_*)} \sum_{m=1}^{M} \sum_{t=t^m_*}^{T^m} [1-|\textbf{u}^{t} - \hat{\textbf{u}}^m|]
    \end{equation*}
    \text{where} ${\textbf{u}}^t$ is the within-tolerance response
    once settled and until the end of the trial $[t^m_*, T^m]$. 
\end{itemize}

 \noindent The outcome measures available immediately after the trajectory following task include: 
\begin{itemize}
    \item \textit{Average stability:} Measured as the dimensionless jerk of the patient trajectory.
    \begin{equation*}
        s = -\frac{T^5}{v^2_{peak}}\int^{T}_0\Big|\frac{d^2v(t)}{dt^2}\Big|^2 dt
    \end{equation*}
    where $v(t)$ is the speed, $T$ is the total trial time, and $v^2_{peak}$ is the maximum of $v(t)$. 
    \item \textit{Average speed:} The average speed during the trajectory following task. 
    \begin{equation*}
        v = \frac{d}{t_{start}-t_{end}}
    \end{equation*}
where $d$ is the Euclidean distance between the start and end position for a straight path segment, and $d$ is the arc length for a curved path segment. $t_{start}$ and $t_{end }$ are respectively the start and end times of a given path segment traversal.
    \item \textit{Percentage of time out of bounds:} The percentage of time during the trajectory following tasks when the simulated wheelchair is outside the indicated path barriers. 
    \begin{equation*}
    t_{ob} = \frac{\sum_{n=1}^{N} t^{n}_i - t^{n}_o}{t^N-t^0}
    \end{equation*}
    where $t^{n}_o$ and $t^{n}_i$ are the $n^{th}$ time the 2-D wheelchair went out of and came back within bounds, respectively, and $N$ is the number of samples in the trajectory.
\end{itemize}

Additionally, the detailed raw data used to calculate the summary statistics for each test condition are stored as an SQL file that can be accessed for further analysis at any time.

\subsection{Practicality, Usability, and Reliability}\label{sec:results}
In terms of practicality, we designed our assessment tool to require minimal equipment which reduces cost, space requirements, and set-up time. The only equipment needed are the person's own electric assistive machine (e.g.,~powered wheelchair), a tablet or any device able to run the assessment application, and our interfacing device (Sec.~\ref{sec:hardware}) if the interface is not already Bluetooth capable. 

In terms of usability, we find that the full assessment can be completed in one session lasting between 30-60 minutes with the default settings, and without taxing the patient or the experimenter. The duration can be adjusted to be longer based on the configurable independent variables, as deemed appropriate by the experimenter. Simple and clear instructions for each task are contained within the app itself, so that the tester can easily administer the assessment. There have been no adverse incidents with the assessment tool as all tasks are simulation-based. The start and end of each test are also clearly defined.


Our motivation was to design an assessment tool that produces outcome measures that are repeatable, consistent, 
precise, and immune to test-taker bias. The outcome measures 
\begin{wrapfigure}[13]{l}{0.206\textwidth}
\centering
\vspace{-.1cm}
\includegraphics[width=\hsize]{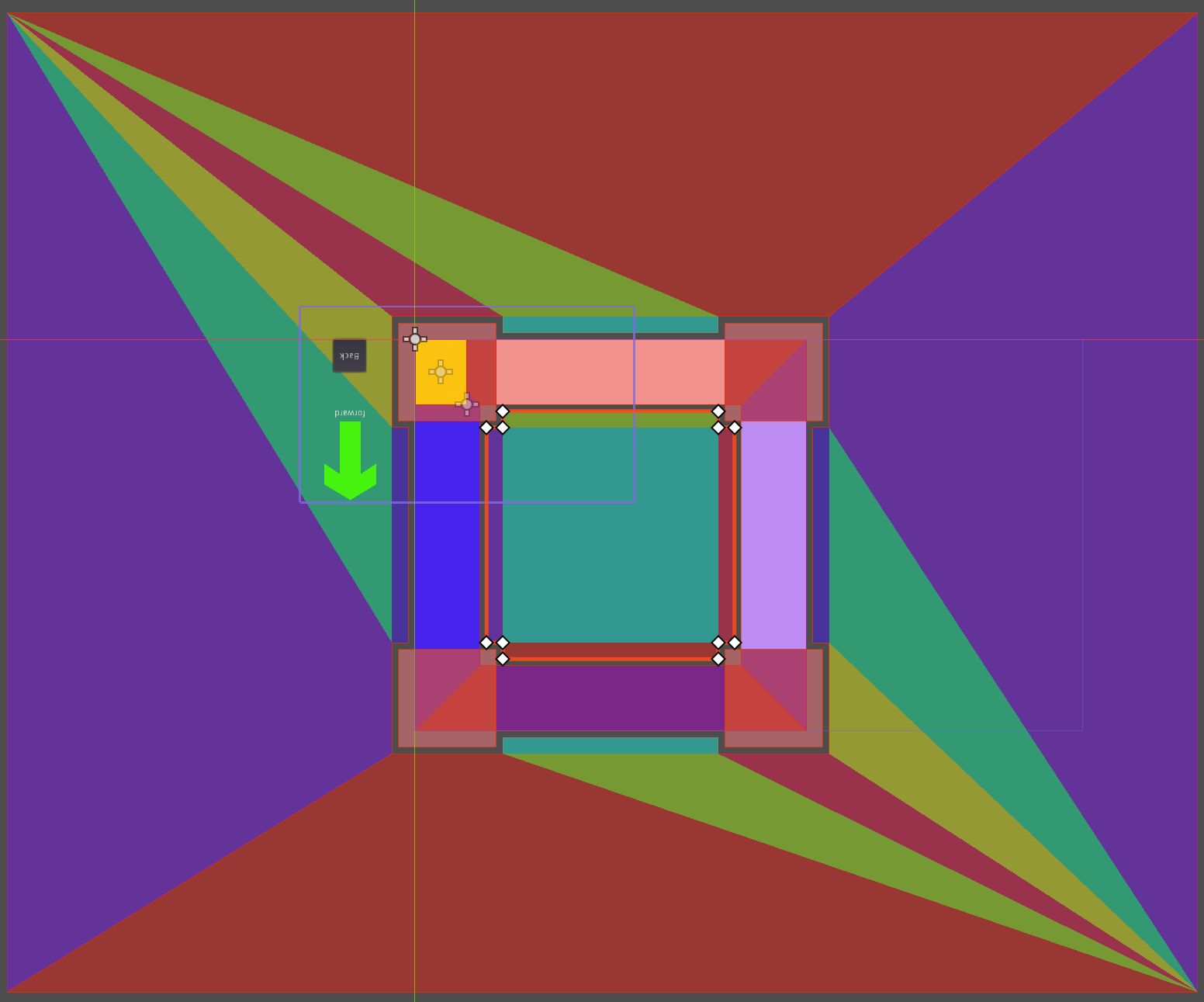}
\caption{Bounding geometries used for calculating out-of-bound duration of the simulated wheelchair.}
\label{fig:boundries}
\end{wrapfigure}
are calculated automatically, which preserves scoring
reliability across various test takers and sessions for a single patient profile. For example, to calculate the total percentage of time the simulated wheelchair is outside of the allowed bounds during the trajectory following task, we use definitive geometries (Fig.~\ref{fig:boundries}) with boundaries designed to account for human field-of-view limitations.


\section{Hardware Connection}\label{sec:hardware}

Some modern powered wheelchairs have Bluetooth-enabled interfaces that allow the interface to connect to digital devices such as computers, smart phones, and tablets. However, many PWs still lack this capability. We have designed a multi-interfacing device that connects to various common interfaces used for the control of powered wheelchairs, such as joysticks, switch-based headarrays, and sip/puff interfaces. Our device serves as a connection that communicates signals from the control interface over Bluetooth, which can then be detected by any device with the interface skills test app. With our open source design, any PW can be used to measure interface usage skill. 

This work was inspired by the Freedom Wing adaptor by AbleGamers~\cite{freedom_wing}. The novel contribution of our work is to replace the wired connection with Bluetooth and to allow for various types of interface connections. 

\begin{figure}
    \centering
    \includegraphics[width=0.47\textwidth]{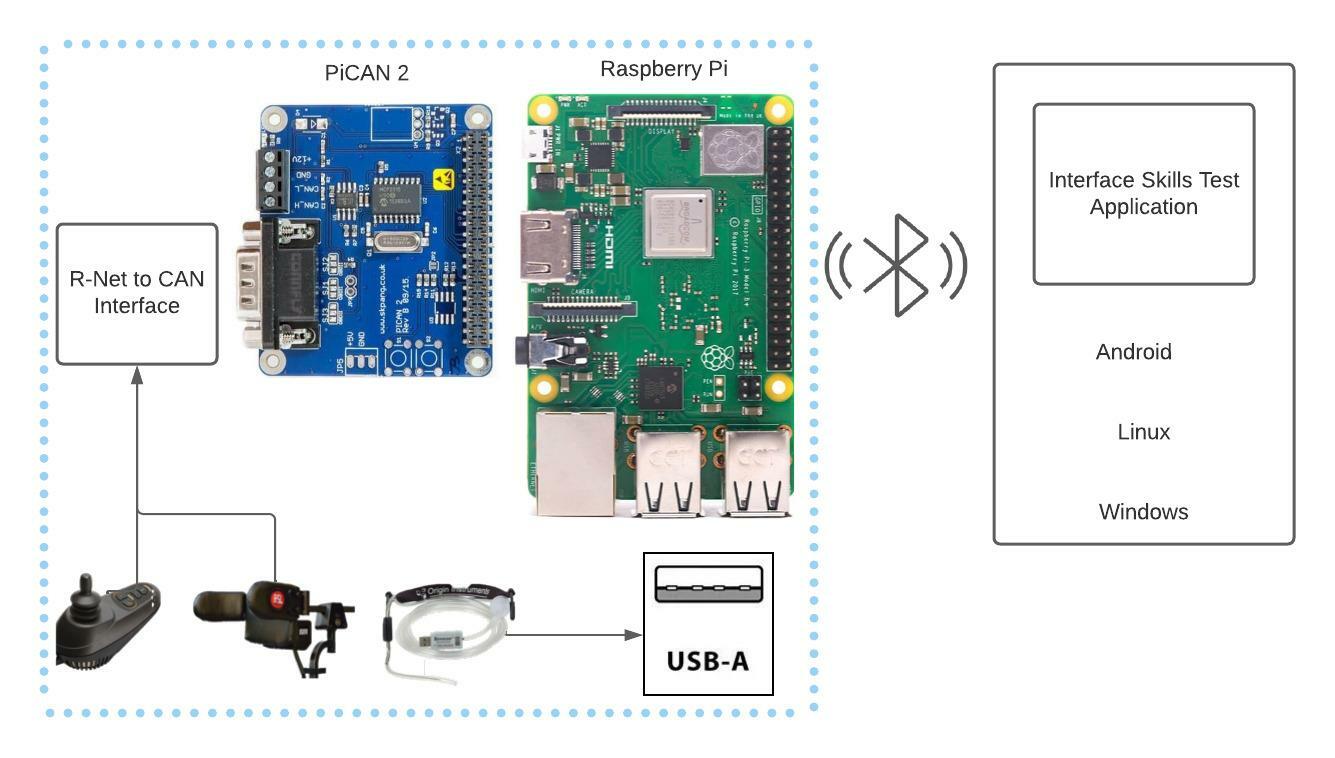}
    \caption{R-Net and USB-A interface to Bluetooth joystick.}
    \vspace{-0.5cm}
    \label{fig:multi_interface}
\end{figure}

The hardware required is minimal and includes off-the-shelf components. A Raspberry Pi\footnotetext[1]{Model B+ was tested.} acts as a bridge by relaying commands from the assistive devices to our app over Bluetooth.\footnotetext[2]{The source code for the device is available at \url{https://github.com/argallab/WheelchairBLEGamepad.git}.} The Raspberry Pi emulates an XBox 360 controller, converting commands from the assistive interface into button presses and joystick values on the controller. An additional PiCAN2 board is used to allow CAN Bus interface connections to the Raspberry Pi, which is required for some R-Net type interface devices. A diagram of the hardware bridge is shown in Figure~\ref{fig:multi_interface}. The hardware connection was tested with R-Net switch-based headarray and joystick interfaces on Permobil M3 and F3 Corpus powered wheelchairs. 

\section{Discussion}\label{sec:discussion}
The goal of designing this assessment tool is to improve the interface training and wheelchair navigation performance by documenting initial and subsequent interface usage skill and characteristics via reliable, repeatable, and objective outcome measures. Reliable and objective measurement instruments are needed not just for providing informed care to patients, but also to assist in testing research hypotheses, comparing outcomes, and developing new technologies. 

The scoring for all tests are digitized and analytical, so the outcome measures are not subject to experimenter bias. 

This assessment also makes possible the study of other interesting research questions. For example, there is potential in using this tool to identify how long-term therapy affects interface usage skill. Also, the tool allows for identifying how various key variables affect different qualities of the human input during interface usage. Furthermore, the tool may aid in deciding which interfaces and which settings are more suitable for a particular individual with evidence-based measurements.  There is also the potential to evaluate how various autonomous robotics assistance interventions affect---either improving or decreasing---the patient's interface usage skill.  

\section{Conclusion and Future Work}\label{sec:conclusion}
In this systems paper, we presented the Interface Skills Test; an assessment tool for evaluating various qualities of interface usage. We anticipate that with automated outcome measure calculations, this assessment tool can aid in understanding a patient's interface usage skill and diagnosing appropriate solutions to overcome deficiencies. This contribution can potentially improve the quality of clinical care and also allow robotics researchers to design customized and intelligent assistance algorithms.

The current version of this assessment tool is limited to 2D assistive machines. In future work, we will expand to cover higher-dimensional machines such as robotic arms. Additionally, our future iteration will include additional tasks and measures to include reachability and operation range. 

We have currently tested the hardware bridge with R-Net controllers. Our next iteration will expand to other common wheelchair controller types. We plan to evaluate test-retest validity, context validity, and usefulness to clinicians via within-subject study.

\section*{ACKNOWLEDGMENT}
This  material  is  based  upon  work  supported  by  the  National  Science  Foundation  under Grant  IIS-1552706. Any opinions, findings, and conclusions or recommendations expressed  in  this  material  are  those  of  the  authors and  do not  necessarily  reflect  the  views  of  the  National  Science Foundation.

\bibliographystyle{unsrt}
\bibliography{references}

\end{document}